\documentclass[aps,pra,floatfix,superscriptaddress,twocolumn]{revtex4-2}
\usepackage{graphicx}
\usepackage{epsfig}
\usepackage{dcolumn}
\usepackage{bm}
\usepackage{braket}
\usepackage{amsmath}
\usepackage{amssymb}
\usepackage{gensymb}
\usepackage{makeidx}
\usepackage{mathtools}
\usepackage{hyperref}
 
\usepackage{hyperref}
\usepackage{multirow}
\usepackage[usenames]{color}

\begin{document}

\title{Out-of-equilibrium dynamics of Bose-Bose mixtures in optical lattices}
\author{Pardeep Kaur}
\affiliation{Department of Physics, Indian Institute of Technology Ropar,
             Rupnagar - 140001, Punjab, India}
\author{Kuldeep Suthar}
\affiliation{Institute of Atomic and Molecular Sciences,
             Academia Sinica, Taipei 10617, Taiwan}
\affiliation{Department of Physics, Central University of Rajasthan,
             Ajmer - 305817, Rajasthan, India}
\author{Dilip Angom}
\affiliation{Department of Physics, Manipur University,
             Canchipur - 795003, Manipur, India}
\author{Sandeep Gautam}
\affiliation{Department of Physics, Indian Institute of Technology Ropar,
             Rupnagar - 140001, Punjab, India}

\date{\today}
   
\begin{abstract}
We examine the quench dynamics across quantum phase transitions from
a Mott insulator (MI) to a superfluid (SF) phase in a two-component bosonic mixture 
in an optical lattice. We show that two-component Bose mixtures exhibit 
qualitatively different quantum dynamics than one-component Bose gas. Besides
second-order MI-SF transitions, we also investigate quench dynamics across a 
first-order MI-SF transition. The Bose mixtures show the critical slowing down 
of dynamics near the critical transition point, as proposed by 
the Kibble-Zurek mechanism. For MI-SF transitions with
homogeneous lattice-site distributions in the MI phase, the dynamical critical
exponents extracted by the power-law scaling of the proposed quantities obtained
via numerical simulations are in very close agreement with the mean-field
predictions.
\end{abstract}

\maketitle

\section{Introduction}
\label{intro}

Ultracold atoms in quantum gas experiments provide an ideal platform to explore
exotic quantum phases and offer unprecedented control over many-body dynamics
that are difficult to probe in condensed matter systems~\cite{UC,UC2}. These
atoms confined within an optical lattice create a captivating and strongly correlated
quantum system~\cite{OL-cold-atoms,MI-SF}. The optical lattice potential is highly 
suitable for simulating many-body Hamiltonians as it facilitates exceptional
tunability of physical parameters, including interparticle interactions,
lattice geometry, particle statistics, lattice depth, filling factors, 
etc., over a broad spectrum. The minimal model that describes the ground-state properties
of bosonic atoms in an optical lattice is the Bose-Hubbard model (BHM)~\cite{BHM}.
The physics underlying BHM holds significance from the perspective of fundamental
understanding and practical applications.

The BHM has underpinned our understanding of quantum phase transitions (QPTs) \cite{QPT,QPT2}. 
The practical relevance of BHM stems from the potential realization of a quantum computer
using a system of cold atoms confined in an optical lattice \cite {QUANTUM-comp}. The study
of out-of-equilibrium dynamics in interacting quantum systems, in search of an adiabatic 
quantum state for quantum computation, is an active research area \cite{qt-comp}.
The controllability of parameters in cold atom systems, especially over time, naturally sparks interest
in the out-of-equilibrium dynamics of the Bose-Hubbard model (BHM), particularly in the proximity 
of quantum critical points. Various methods can be employed to take a quantum system out of equilibrium, 
such as connecting it to an external bath or applying a driving field. A straightforward approach
involves modifying one of the parameters in the underlying Hamiltonian
of the system termed a quantum quench. The Kibble-Zurek \cite{Kibble, Zurek} mechanism (KZM)
offers a comprehensive theoretical framework to understand the non-equilibrium dynamics of such systems and 
predicts a universal power-law scaling of excitations in relation to the quench rate, with an exponent directly
related to the equilibrium critical exponents \cite{Kibble,Zurek,Kibble-2,Zurek-2}. Initially 
proposed to explain the evolution of the early universe \cite{Kibble}, the KZM has been experimentally 
explored in various classical and quantum phase transitions. Its application has been demonstrated in 
numerous systems such as cosmic microwave backgrounds \cite{Cos-WG}, liquid helium \cite{L-He}, 
superconductors \cite{Super-cond}, and liquid crystals \cite{Liquid-Cry}. More recently, the studies on KZM 
have been extended to ultracold quantum gases \cite{KZM-UC-atoms-expt-BEC,KZM-UC-BHM-EXPT,KZM-FERMI-SF}, 
where theoretical \cite{Shimizu-1,Shimizu-2, Reijish, Shimiju3, hrushi, cavity} and experimental studies 
\cite{KZM-UC-BHM-EXPT,bhm-expt2,bhm-expt3} till now have focused on the one component Bose-Hubbard model. On the 
other hand, the quench dynamics of a mixture of two-component bosonic systems have not been investigated yet.

In the present work, we theoretically examine the quench dynamics of Mott insulator-superfluid (MI-SF) 
phase transitions in the two-component Bose-Hubbard model. A two-component Bose mixture in an optical 
lattice is not just an extension of a one-component Bose gas in a lattice as the phases of multi-component and 
spinor systems in such scenarios are considerably more intricate \cite{Two-comp-BHM}. The dynamics of two-component
BHM differ from the corresponding MI-SF transition in one-component BHM. 
The distinction arises due to the inhomogeneity of the MI and SF phases of the mixture. In the miscible
domain, both components can occupy the same lattice site in an even-integer Mott lobe and the superfluid phase.
This leads to homogeneous atomic occupancy distributions of the two components,
whereas for the odd-integer Mott lobes, the atomic occupancy distributions are inhomogeneous. In the
phase-separated domain, due to stronger interspecies repulsion, it is not possible for both components to occupy 
the same lattice site in MI and SF phases. Hence, the qualitative features of the MI-SF phase transitions in the 
two-component system are different leading to novel quantum dynamics. Moreover, the MI-SF transition in the 
one-component BHM is a continuous transition, whereas the two-component BHM exhibits a tricritical point after
which the MI-SF transition changes to a first-order transition \cite{Tricritical}. Although KZM predicts the 
breakdown of adiabaticity for continuous transitions, both experimental \cite{1st-order-expt1, 1st-order-expt2} 
and theoretical studies \cite{Shimiju3,hrushi,1st-order} confirm the critical slowing down for the first-order phase
transitions also. Motivated by these studies, we explore both the first-order and second-order MI-SF transitions
of the two-component BHM in the present work. We have discussed the effects of the inhomogeneity of phases and order
of transitions on the impulse regime and scaling 
exponents.

This paper is organized as follows. In section~\ref{statics}, we introduce the two-component
Bose-Hubbard model and describe the mean-field approach to determine the equilibrium
phase diagram for three different values of the inter-species interactions corresponding to the
miscible-immiscible phase transition. The dynamical Gutzwiller equations and the
Kibble-Zurek mechanism is presented in section~\ref{dynamics}. Section~\ref{res} discusses
the quantum quench dynamics across MI(2)-SF and MI(1)-SF transitions of two-component BHM. 
Finally, we summarize our findings in section ~\ref{summary}.


\section{Two-component Bose-Hubbard Model}
\label{statics}
We consider a Bose-Bose mixture of two spin states from the same hyperfine-spin manifold in a two-dimensional 
(2D) square optical lattice at zero temperature. The two-component Bose-Hubbard model (TBHM) describes a 
mixture of bosonic species. The model Hamiltonian\cite{TBHM} is
\begin{eqnarray}
\hat{H} &=& -\sum_{p,q,\sigma}\bigg [ \Big( J_x^{\sigma} 
              \hat{b}_{p+1, q}^{\dagger \sigma}\hat{b}_{p,q}^\sigma + {\rm H.c.}\Big)
              + \Big( J_y^\sigma \hat{b}_{p,q+1}^{\dagger \sigma}
              \hat{b}_{p,q}^\sigma  \nonumber\\ 
              &&+ {\rm H.c.}\Big) - \frac{U_{\sigma\sigma}}{2}
              \hat{n}_{p,q}^\sigma (\hat{n}_{p,q}^\sigma-1) + 
              {\mu}_{\sigma}\hat{n}_{p,q}^\sigma\bigg]
              \nonumber\\
              &&+\sum_{p,q} U_{\uparrow\downarrow}\hat{n}_{p,q}^{\uparrow} \hat{n}_{p, q}^\downarrow,
\label{tbhm}  
\end{eqnarray}
where $(p,q)$ is the index of lattice sites with $p$ and $q$ as the indices 
along $x$- and $y$-directions, respectively. Here $\sigma = (\uparrow,\downarrow)$ is the spin-state index,
$J_x^\sigma$ ($J_y^\sigma$) is the nearest-neighbour hopping strength along $x$ ($y$) direction,
$\hat{b}^{\dagger \sigma}_{p,q}$ ($\hat{b}^\sigma_{p,q}$) is the creation (annihilation) operator, 
$\hat{n}_{p,q}^\sigma$ is the number operator at the site ($p,q$), $U_{\sigma\sigma}$ is intraspecies
interaction strength, and $U_{\uparrow\downarrow}$ is the interspecies interaction 
strength between the two components, and ${\mu}_{\sigma}$ is the chemical potential of $\sigma$ spin-component.

\subsection{Static Gutzwiller Mean-Field Theory}
\label{staticsA}
To examine the ground-state properties of the model Hamiltonian
[Eq.~\ref{tbhm}], we use the single-site Gutzwiller mean-field (SGMF) theory \cite{SGMF}. We decompose annihilation
($\hat{b}^\sigma_{p,q}$) and creation ($\hat{b}^{\dagger \sigma}_{p,q}$) operators into the mean-field and fluctuation
part as 
$\hat{b}_{p, q}^\sigma = \phi_{p,q}^\sigma + \delta \hat{b}_{p, q}^\sigma$ and $\hat{b}^{\dagger \sigma}_{p,q} 
= \phi_{p,q}^{\sigma*} + \delta\hat{b}_{p,q}^{\dagger \sigma}$,
where $\phi_{p,q}^\sigma$ ($\phi_{p,q}^{\sigma*}$) is the superfluid order parameter. The above approximation
decouples the Hamiltonian [Eq.~(\ref{tbhm})] in lattice sites, and the Hamiltonian can be written as the sum
of single-site Hamiltonians. The local Hamiltonian at the $(p,q)$th site is
\begin{eqnarray}
\hat{h}_{p,q} &= &- \sum_{\sigma}\left[J_x^\sigma 
                  \left(\hat{b}_{p+1, q}^{\dagger \sigma}\phi_{p,q}^\sigma 
                  + \phi^{*\sigma}_{p + 1, q}\hat{b}_{p, q}^\sigma \right ) + {\rm H.c.}
                  \right.  \nonumber\\
                  && +  J_y^\sigma\left(\hat{b}_{p, q+1}^{\dagger \sigma} \phi_{p,q}^\sigma 
                  +  \phi^{*\sigma}_{p, q+1}\hat{b}_{p, q}^\sigma  \right) + {\rm H.c.}
                  \nonumber \\
                  &&-\left . \frac{U_{\sigma\sigma}}{2}\hat{n}_{p, q}^\sigma
                  \left (\hat{n}_{p, q}^\sigma-1\right) 
                  + {\mu}_{\sigma}\hat{n}_{p, q}^\sigma\right]
                  + U_{\uparrow\downarrow}\hat{n}_{p, q}^\uparrow \hat{n}_{p, q}^\downarrow.
                  \nonumber\\
\label{ham_ss_tbec}
\end{eqnarray}
The total mean-field Hamiltonian of the system is $\hat{H} = \sum_{p,q}\hat{h}_{p,q}$. To obtain the
ground state, we self-consistently diagonalize the Hamiltonian at each lattice site. The many-body
Gutzwiller wave function for the ground-state at the $(p,q)$th site is~\cite{ansatz}
\begin{equation}
  |\Psi\rangle = \prod_{p,q}|\psi\rangle_{p,q} = \prod_{p,q}\sum_{n_{\uparrow}, n_\downarrow }c^{(p,q)}_{n_\uparrow, n_\downarrow} 
                       |n_\uparrow, n_\downarrow\rangle_{p,q}.
  \label{gw_2s}
\end{equation}
Here $|n_{\uparrow}\rangle$ and $|n_\downarrow\rangle$ are Fock states with  
$n_\sigma \in [0,N_b-1]$, where  $N_b$ is the dimension of the Fock space. The c-numbers $c_{n_{\uparrow},n_\downarrow}^{p,q}$'s 
are the complex coefficients that satisfy the normalization condition
$\sum_{n_{\uparrow},n_{\downarrow}}
|c^{(p,q)}_{n_{\uparrow},n_{\downarrow}}|^2$ = 1. The superfluid order parameters of the two components are 
\begin{subequations}
  \begin{eqnarray}
    \phi_{p,q}^{\uparrow}& =& _{p,q}\langle\psi|\hat{b}_{p,q}^\uparrow|\psi\rangle_{p,q} 
            = \sum_{n_\uparrow, n_\downarrow}\sqrt{n_\uparrow} 
              {c^{(p,q)*}_{n_\uparrow-1, n_\downarrow}}c^{(p,q)}_{n_\uparrow,n_\downarrow},
              \label{gw_phi2s_1}\\
    \phi_{p,q}^\downarrow& =& _{p,q}\langle\psi|\hat{b}_{p,q}^\downarrow|\psi\rangle_{p,q} 
            = \sum_{n_\uparrow, n_\downarrow}\sqrt{n_\downarrow} 
              {c^{(p,q)*}_{n_\uparrow, n_\downarrow-1}}c^{(p,q)}_{n_\uparrow,n_\downarrow}.
    \label{gw_phi2s_2}              
  \end{eqnarray}
\end{subequations}
The atomic occupancies of the components at a lattice site $(p,q)$ are the expectation of the number
operators and are given by
\begin{subequations}
  \begin{eqnarray}
   \rho_{p, q}^\uparrow& =& _{p,q}\langle\psi|\hat{n}_{p, q}^\uparrow|\psi\rangle_{p,q} 
            = \sum_{n_\uparrow, n_\downarrow} n_\uparrow |c^{(p,q)}_{n_\uparrow,n_\downarrow}|^2,
            \label{num2s_1}\\
   \rho_{p, q}^\downarrow& =& _{p,q}\langle\psi|\hat{n}_{p, q}^\downarrow|\psi\rangle_{p,q} 
            = \sum_{n_\uparrow, n_\downarrow}n_\downarrow|c^{(p,q)}_{n_\uparrow,n_\downarrow}|^2. 
   \label{num2s_2}              
  \end{eqnarray}
\end{subequations}
Since the local Hamiltonian $\hat{h}_{p,q}$ depends on $\phi_{p,q}^\sigma$ and $\hat{n}_{p,q}^\sigma$, 
therefore in our numerical computations the initial state is considered as a complex random distribution
of the Gutzwiller coefficients across the lattice. We, then, solve for the ground state of each site by
diagonalizing the corresponding single-site Hamiltonian. Using the updated superfluid order parameters,
the ground state of the next lattice site is computed. This process is repeated until all the sites
of a square lattice are covered. One such sweep is identified as an iteration, and we start the process
again for the next iteration. The iterations are repeated until the requisite convergence criteria 
are satisfied. In our study, to compute the equilibrium phase diagrams, we used $50$ initial
random configurations. This is to ensure that the minimum energy state has been achieved. We consider a lattice
size of $8\times 8$ and $N_b = 6$, and checked that by increasing the system size 
or $N_b$, phase diagrams do not alter.

\subsection{Equilibrium Phase Diagrams}
\label{staticsB}
In the present work, we consider equal hopping strengths in both directions, i.e. $J_x^{\sigma} = J_y^{\sigma} = J$, 
and identical chemical potentials ${\mu}_{\sigma} = \mu$ and intraspecies interactions $U_{\sigma\sigma} = U$.
We scale all energies with respect to $U$. Following the phase separation criteria of the two components, determined
by the strength of the interspecies interaction, we investigate the two regimes: $U_{\uparrow\downarrow} < 1$ 
and $U_{\uparrow\downarrow} > 1$. We use the Gutzwiller mean-field approach to obtain phase diagrams of TBHM.
The model exhibits two phases: MI and SF phases~\cite{bai,Ozaki-arxiv,MF-tri,MC-tri}.
\subsubsection{Interspecies Interaction $U_{\uparrow\downarrow} < 1$ }
\label{staticsB-1}
\begin{figure}[h!]
  \includegraphics[width=\linewidth]{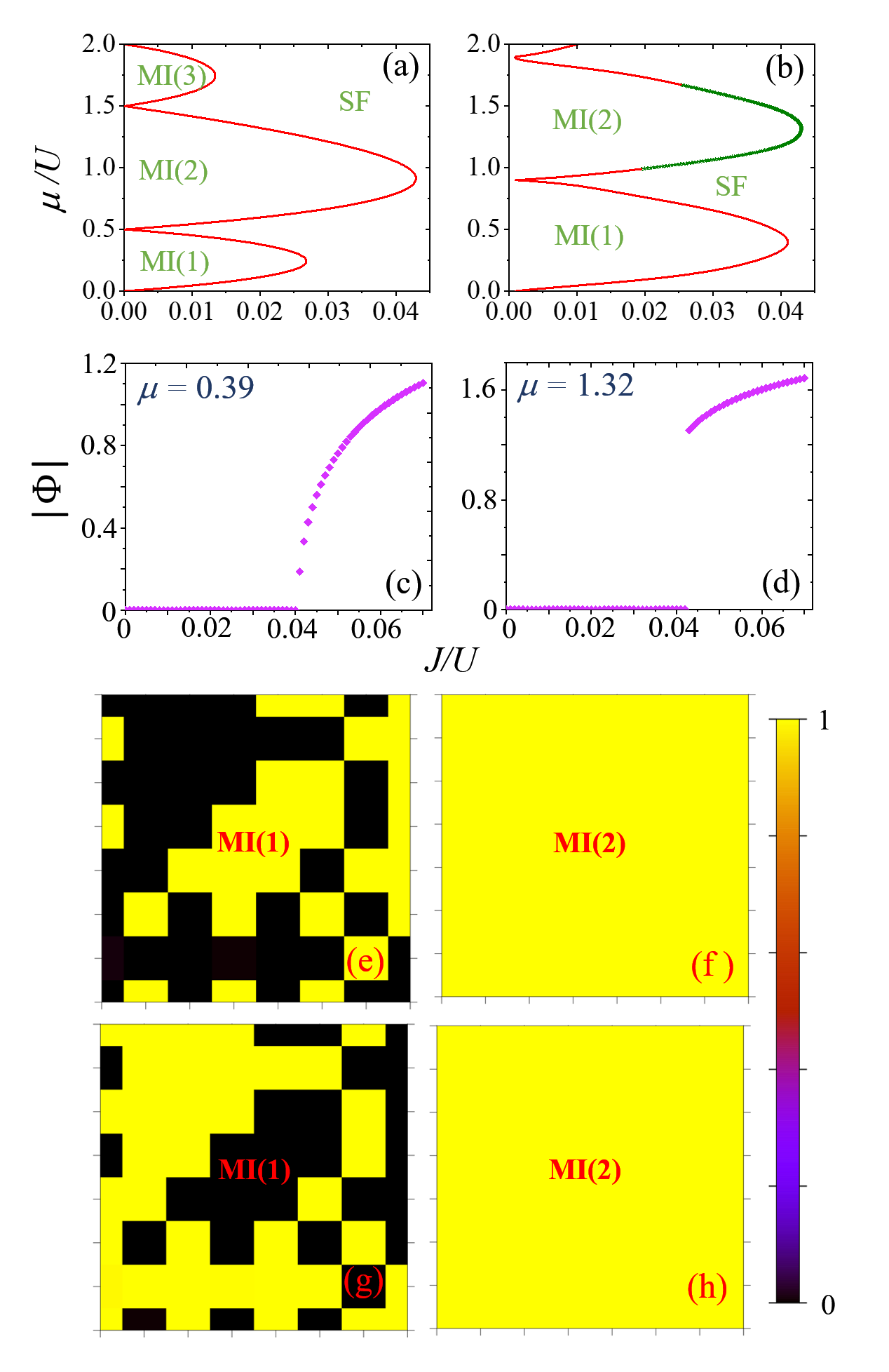}
   \caption{Phase diagrams of the two-component BHM in $J$-$\mu$ plane for (a) $U_{\uparrow\downarrow} = 0.5$ and 
    (b) $U_{\uparrow\downarrow} = 0.9$. The number in parentheses is the total average atomic occupancy.
     The order of phase transition across all boundaries is second order except for the green curve in (b) 
     across which the MI(2) to SF transition is of the first order in nature. 
    For $U_{\uparrow\downarrow}=0.9$, the $|\Phi|$ as a function of $J$ are shown for 
    (c) MI(1)-SF transition at $\mu=0.39$ and (d) MI(2)-SF transition at $\mu=1.32$.  
    In the third and fourth panels [e-h], sample atomic occupancy distributions ($\rho_{p,q}^\sigma$) of both components 
    on an $8\times8$ square lattice are shown. 
    (e) and (g) correspond to $\rho_{p,q}^\uparrow$ and $\rho_{p,q}^\downarrow$, respectively, in the MI(1) phase. The same
    for MI(2) phase are in (f) and (h).} 
\label{FIG1}
\end{figure}

We first consider $U_{\uparrow\downarrow} = 0.5$ and $0.9$ in the miscible regime of TBHM. In the MI phase,
$\phi^{\sigma}_{p,q}$ is zero, but it is nonzero in the superfluid phase. We use this criterion to determine
the phase boundary between MI and SF phases in the $J$-$\mu$ plane. Figs. \ref{FIG1}(a) and (b) show the 
phase diagrams for $U_{\uparrow\downarrow} = 0.5$ and $0.9$.

The phase diagram depicts two kinds of MI lobes based on  average total occupancy. These are even-integer and odd-integer
MI lobes. In $J=0$ limit, the size of the MI regions on the $\mu$-axis for odd and even total fillings are 
$U_{\uparrow\downarrow}$ and $1$, respectively. As we increase $U_{\uparrow\downarrow}$, the sizes of odd Mott lobes along $J$-axis 
increases while those of even Mott lobes remain the same. 
For MI(1) phase, the average occupancies $\rho^{\sigma} 
= \sum_{p,q}\rho^\sigma_{p,q}/N_{\rm lattice}$ are $0 <\rho^{\uparrow} <1$, and $\rho^{\downarrow} = 1-\rho^{\uparrow}$, and for 
the even-integer MI(2) phase, $\rho^{\uparrow} = \rho^{\downarrow} = 1$ \cite{bai}.

At $U_{\uparrow\downarrow}=0.5$, the MI-SF quantum phase transitions for both odd and even-occupancy Mott lobes are second-order 
transitions. However, at $U_{\uparrow\downarrow}=0.9$ the MI-SF transitions are not entirely second order. We find 
that the change in the order-of-transition, near the tip of the MI(2) lobe, occurs at $U_{\uparrow\downarrow}=0.65$. For 
$U_{\uparrow\downarrow} = 0.9$, as shown in Fig.~\ref{FIG1}(b), the tricritical points on $\mu$-axis exist at $0.99$ and 
$1.67$~\cite{Tricritical, Ozaki-arxiv,MF-tri}. For the regime between these two $\mu$ values, the MI(2)-SF transition is of 
first order and marked by green points in Fig. \ref{FIG1}(b). 
To confirm the order of these transitions, we have calculated the amplitude of the superfluid order parameter as a function of $J$ for
a fixed $\mu$ across the critical hopping strength. 
The continuous variation of $|\Phi| = \sum_{p,q,\sigma} |\phi^\sigma_{p,q}|/N_{\rm lattice}$ with $J$ represents 
a second-order phase transition as illustrated in Fig. \ref{FIG1}(c) 
for a MI(1)-SF phase transition with $\mu = 0.39$ 
and $U_{\uparrow\downarrow}=0.9$, here $N_{\rm lattice}$ is the number of lattice sites. 
On the other hand, a discontinuity in $|\Phi|$ across  MI(2)-SF phase boundary
in Fig.~\ref{FIG1}(d) for $\mu = 1.32$ and $U_{\uparrow\downarrow}=0.9$ is indicative
of the first-order phase transition.

The sample atomic occupancy distributions $\rho^{\sigma}_{p,q}$ in MI(1) and MI(2) phases confined in $N_{\rm lattice} = 8\times8$ 
square lattice are shown in Figs.~\ref{FIG1} (e),(g) and Figs.~\ref{FIG1} (f),(h), respectively. 
The atomic occupancy distributions in the SF phase are uniform and identical for both the components 
with real average occupancy.

\subsubsection{Interspecies Interaction $U_{\uparrow\downarrow} > 1$}
\label{staticsB-2}
\begin{figure}[ht]
  \includegraphics[width=\linewidth]{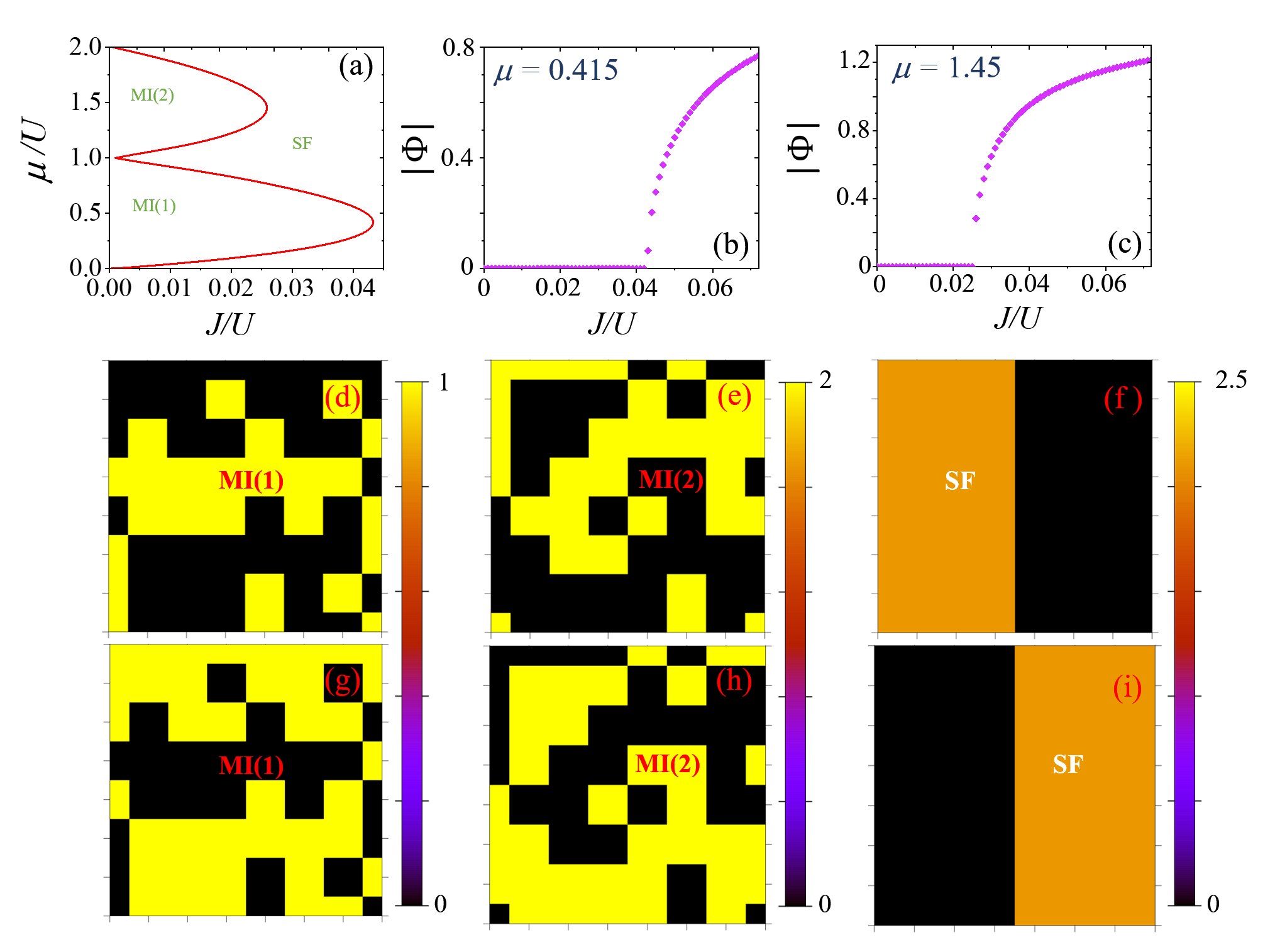}
   \caption{ (a) The phase diagram of the TBHM in $J-\mu$ plane in the immiscible regime. 
   (b) The variation of $|\Phi|$  as a function of $J$ with $\mu = 0.415$, corresponding to MI(1)-SF phase transition 
    and (c) with $\mu = 1.45$ corresponding to MI(2)-SF transition; these are plotted for 
   $U_{\uparrow\downarrow}=1.5$. The sample $\rho_{p,q}^\uparrow$ distributions in (d) MI(1), (e) MI(2), and (f) SF phases for 
   $U_{\uparrow\downarrow}=1.5$. 
   Similarly, corresponding $\rho_{p,q}^\downarrow$ distributions are in (g), (h), and (i).}
   \label{Fig2}
\end{figure}

For $U_{\uparrow\downarrow} > 1$, phase separation of the mixture of bosonic species occurs. 
We have shown the phase diagram for $U_{\uparrow\downarrow}=1.5$ in Fig.~\ref{Fig2}(a) which does not
change with an increase in $U_{\uparrow\downarrow}$.
The continuous nature of the amplitude of the average superfluid order parameter $\Phi$ as a function of $J$ for $\mu = 0.415$ and 
$1.45$ in Figure~\ref{Fig2}(b) and Figure~\ref{Fig2}(c), respectively, confirms the second-order nature of the MI-SF phase transitions.
In this regime, only one of the components, chosen randomly, occupies the lattice site. 
The occupancy of the other component remains zero at that site. This applies to both MI and SF phases. 
However, for the latter, the occupancy of one component is a real number, and for the other is zero. 

 The sample atomic occupancy distributions for MI(1), MI(2), and SF phases are presented in Fig~\ref{Fig2} (d),(g), 
 Fig~\ref{Fig2} (e),(h), and Fig~\ref{Fig2} (f),(i), respectively.

\section{Quench Dynamics}
\label{dynamics}
\subsection{Time-dependent Gutzwiller Approach}
To study the out-of-equilibrium dynamics, we quench the hopping parameter in time, $J \xrightarrow{} J(t)$ whereas 
all other model parameters remain steady in time. This results in the time dependence of TBHM. As $J$ is ramped up 
in quench, the system is expected to undergo MI-SF phase transition. The time evolution of the single-site Gutzwiller
wave function is governed by the time-dependent Gutzwiller equation given as
\begin{equation}
   i\hbar\partial_t |{\psi}\rangle_{p,q} = \hat{h}_{p,q} |{\psi}\rangle_{p,q}.
   \label{schdinger_eqn}
\end{equation}
This leads to a system of coupled linear partial differential equations for the coefficients $c^{(p,q)}_{n_\uparrow,n_\downarrow}(t)$. 
To solve the coupled equations, we employ the fourth-order Runge-Kutta method. In this method, we first obtain the  
coefficients $c^{(p,q)}_{n_\uparrow,n_\downarrow}(t)$ using the static Gutzwiller approach, 
and then the wave function of the system at a specific 
time instant $t$ is computed. However, to derive the quantum phase transition, quantum fluctuations are needed.
To generate the effects of quantum fluctuations, we add an initial random noise to the equilibrium coefficients
$c^{(p,q)}_{n_\uparrow,n_\downarrow}$ of the state at the start of
the quench. We first generate univariate random phases within the range of $[0, 2\pi]$ and add them to the
non-zero coefficients. Next, we add density fluctuations by applying noise to the amplitudes of the coefficients. 
This is achieved by generating univariate random numbers 
within the range of $[0,\delta]$, where $\delta$ is set to be $10^{-4}$ in the 
present study. To ensure reliable results, we consider $10$ initial states that are randomized using the methods 
described previously. Each of these initial states is then evolved in time by performing the appropriate parameter 
quench. We further calculate the physical observables of interest averaged over all these 10 samples. Additionally,
for each sample, the observable is averaged across the entire lattice.

\subsection{Kibble-Zurek Mechanism}
\label{dynamicsA}
In the present work, we consider a linear quench protocol given as 
\begin{equation}
  J(t) = J_i + \frac{(J_c - J_i)}{\tau_Q}(t + \tau_Q). \quad t \in [-\tau_Q,t],
\end{equation}
which involves a quench time $\tau_Q$ to determine the quench rate. The critical value of $J$, denoted by $J_c$, is 
crossed at $t=0$. It is assumed that the quench is initiated at time $t=-\tau_Q$, such that 
$J(-\tau_Q)=J_i$. The system's relaxation time determines the fate of the evolving state. 

When the quenched parameter is far from the critical point, the relaxation time is small. 
This results in an adiabatic evolution where the evolving state is close to the actual 
ground state. However, as the critical point is approached, the divergence of relaxation 
time breaks the adiabaticity, leading to a frozen state. This state does not change for some time near the critical point.
The time interval during which the state remains frozen is termed the impulse regime. It starts evolving after a time 
instant $\hat{t}$ that is delayed time or transition time after which evolution is again adiabatic. Thus, the two 
adiabatic regimes are separated by a non-adiabatic regime near the critical point. The transition between these 
regimes is a significant aspect of KZM. The non-adiabatic evolution of the system during the quench inevitably 
leads to excitations and defects in the evolved state. 

In the Kibble-Zurek hypothesis, for a second-order phase transition, the scaling relation between the quench time 
$\tau_Q$ and the transition time $\hat{t}$ is defined by $\hat{t}\propto \tau_{Q}^{\nu z/(1+\nu z)}$, where $\nu$ 
is the critical exponent of the equilibrium correlation length and $z$ is the dynamical critical exponent. 
Additionally, the scaling relation between the density of defects ($N_d$) and $\tau_Q$ in two dimensions is
given by $N_d \propto \tau_{Q}^{-2\nu/(1+\nu z)}$. These relations predict the critical behavior of the 
system near the phase transition and the formation of topological defects during the non-adiabatic 
evolution of the system. In our case, i.e. during the transition from MI to the SF phase, the global
$U(1)$ symmetry spontaneously breaks and gives rise to the vortices. The density of vortices in an
optical lattice system can be computed as \cite{Shimizu-1,Shimizu-2,Reijish,hrushi}
\begin{equation}
  N_{v}^\sigma = \sum_{p,q} |\Omega_{p,q}^\sigma|, 
  \label{vort_den}
\end{equation}
with
\begin{eqnarray}
  \!\!\!\!\!\!\!\!
  \Omega_{p,q}^\sigma &=& \frac{1}{4}\big [\sin(\theta_{p+1,q}^\sigma - \theta_{p,q}^\sigma)
                   + \sin(\theta_{p+1,q+1}^\sigma - \theta_{p+1,q}^\sigma)  
                                         \nonumber \\
               &&  -\sin(\theta_{p+1,q+1}^\sigma - \theta_{p,q+1}^\sigma) - 
                   \sin(\theta_{p,q+1}^\sigma - \theta_{p,q}^\sigma)\big].
  \label{vort_def}
\end{eqnarray}
Here, $\theta_{p,q}^\sigma$ is the phase of the SF order parameter $\phi_{p,q}^\sigma$.
Another quantity that serves as an analog to the defect density is the excess energy above the ground state. 
This quantity is termed residual energy~\cite{residual, residual1, residual2}. This residual energy
$E_{\rm res}$ is given by $E_{\rm res} = E_{\rm fin} - E_{\rm gs}$, where 
$E_{\rm fin} = \langle\Psi(t)|\hat{H}(t)|\Psi(t)\rangle$ denotes the energy of the system at time $t$, 
while $E_{gs}=\langle\Psi_{gs}(t)|\hat{H}(t)|\Psi_{gs}(t)\rangle$ is the ground-state energy for 
Hamiltonian at time $t$. Slower is the evolution, smaller is the residual energy. The scaling relation 
for residual energy is $E_{\rm res} \propto -\tau_{Q}^{2\nu/(1+\nu z)}$ \cite{residual, residual1, residual2}. 
The scaling relations are valid at $\hat{t}$ but it is very difficult to estimate $\hat{t}$ from numerical
simulations. Prior works are based on determining $\hat{t}$ based on $\Phi$~\cite{Shimizu-1,Shimizu-2,Reijish,Shimiju3,cavity}. 
The growth time of the superfluid order parameter depends on the amount of random fluctuations as pointed out in 
Ref. \cite{machida-IKZM}. We choose the protocol to determine $\hat{t}$ used in  Ref.\cite{hrushi}. We calculate
the overlap $O(t) = |\langle{\Psi(0)}|{\Psi(t)}\rangle|$. Since the dynamics are frozen in the impulse regime,
$O(t)$ would be equal to unity until the state is in the impulse regime, as soon as it deviates from unity, 
it indicates that the adiabatic regime has begun and that time instant is $\hat{t}$. The observables 
$\Phi$, $N_v^\sigma$, $E_{\rm res}$, and $O(t)$ relevant to quench dynamics are obtained by averaging over 
10 initial states perturbed by different random noise distributions. 

\section{Results and Discussions}
\label{res}
\subsection{$U_{\uparrow\downarrow} = 0.5$}
\label{dynamicsB}
\subsubsection{MI(2) to SF phase transition}
\label{dynamicsB-1}
\begin{figure}[ht]
  \includegraphics[width=0.8\linewidth]{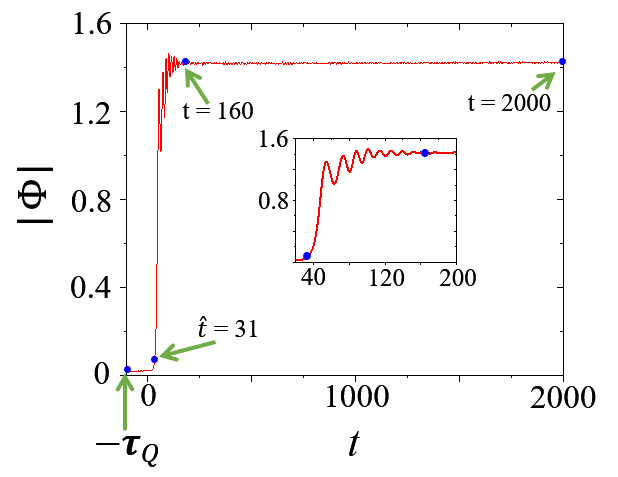}
  \caption{The time evolution of $|\Phi|$ for MI(2)-SF transition with $U_{\uparrow\downarrow}=0.5$, $\mu = 1$, 
            and $\tau_Q = 100$.
            Blue points indicate temporal markers referred to in the main text. In the inset, we show the 
            enlarged view of the dynamical evolution at 
           shorter times from $t=20$ to $t=200$. $|\Phi|$ is nearly zero for $t<\hat{t} = 31$.}
\label{plot-phi}
\end{figure}

Considering $\mu=1$, we start a quench of the hopping parameter from $J_i=0.02$ that 
lies deep within the MI lobe. We end the quench at $J_f=0.064$ within the SF phase.
We confirm the slowing down of transition from the growth of the superfluid order 
parameter that starts after the critical value $J_c$ is passed. We have 
shown one such dynamics for $\tau_Q=100$ in Fig.~\ref{plot-phi}. $|\Phi|$ is close to 
zero until $J$ passes the critical $J_c$ at $t=0$. After 
$t= \hat{t}=31$, $|\Phi|$ shows a sudden increase followed by rapid oscillations. 
Quench is stopped at $t=\tau_Q = 100$. At a longer time, the $|\Phi|$ stabilizes 
after small oscillatory transients. For illustration, the stable $|\Phi|$ is 
marked at $t=160$ in Fig.~\ref{plot-phi}.
Due to the random noise added to the initial Gutzwiller coefficients, there are many vortices at the beginning of 
the quench. During MI to SF phase transition, when the system enters the SF phase, one expects a coherent phase
throughout the system. This is due to the breaking of $U(1)$ global gauge symmetry. However, the quench dynamics
leads to the formation of domains in the system which indicates the existence of local choices of broken symmetry 
in SF phase. This results in a domain structure as predicted by the KZM. The phase singularities at the domain boundaries 
correspond to the vortices, as confirmed by the phase variations. As the system enters into the deep SF phase, the
size of domains increases through domain merging. This results in a decrease in the number of topological defects due to 
pair annihilation, and the system attains phase coherence at long-time evolution.
\begin{figure}[h]
  \includegraphics[width=\linewidth]{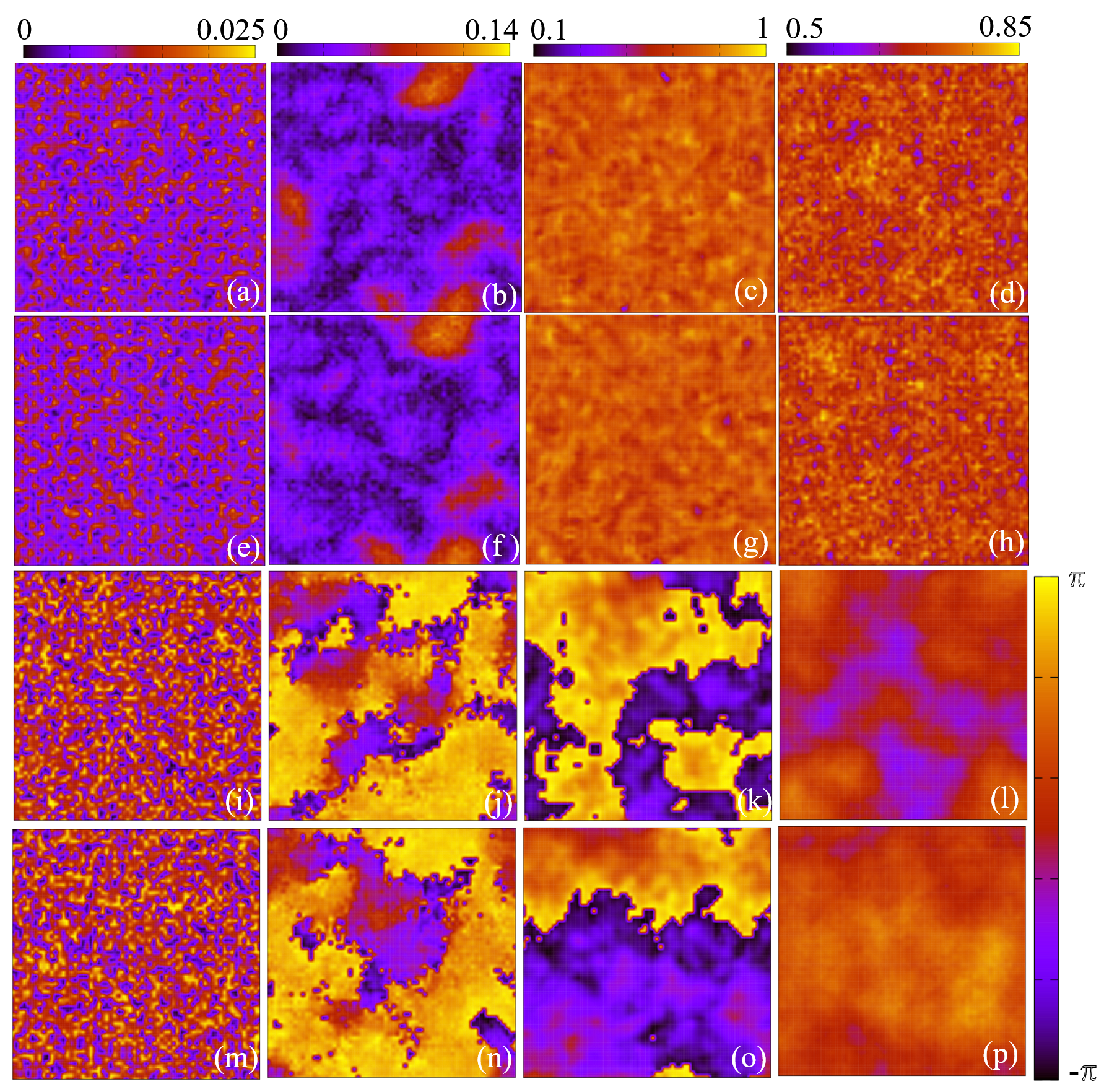}
  \caption{$|\phi^\uparrow_{p,q}|$ at (a) $t=-100$, (b) $t= 31$, (c) $t = 160$, and (d) $t = 2000$ for MI(2)-SF
  phase transitions corresponding to the time evolution in Fig. \ref{plot-phi}. Similarly, $|\phi^\downarrow_{p,q}|$ 
  at the same instants are in (e)-(h). Phases corresponding to $|\phi^\uparrow_{p,q}|$ in (a)-(d) are shown in (i)-(l),
  and the same for  $|\phi^\downarrow_{p,q}|$ in (e)-(h) are in (m)-(p).} 
\label{instant-up5}
\end{figure}

To illustrate the domain formation and merging in the quench dynamics starting with a single randomized initial state at 
$t = -\tau_Q = -100$, we have presented the snapshots of $|\phi^{\sigma}_{p,q}(t)|$ and the respective phases at various
times. At the beginning of the quench the $|\phi^\uparrow_{p,q}|$ and $|\phi^\downarrow_{p,q}|$ have random lattice-site 
distributions with peak values nearly zero as shown in Fig.~\ref{instant-up5}(a) and 
\ref{instant-up5}(e), respectively. The phases of the order parameters for both components are also random, as shown in 
\ref{instant-up5}(i) and \ref{instant-up5}(m), respectively. As time evolves, at $t = \hat{t}=31$, $|\phi^\uparrow_{p,q}|$ 
and $|\phi^\downarrow_{p,q}|$ acquire relatively large peak values accompanied by domain formation. This is presented in 
\ref{instant-up5}(b) and \ref{instant-up5}(f); the corresponding phases are illustrated in Figs. \ref{instant-up5}(j) 
and \ref{instant-up5}(n). The hopping parameter quenching is stopped at $t=100$ as stated earlier. At $t=160$, 
$|\phi^\uparrow_{p,q}|$ and $|\phi^\downarrow_{p,q}|$ acquire almost uniform distributions as shown in Fig. \ref{instant-up5}(c)
and \ref{instant-up5}(g), and their respective phases \ref{instant-up5}(k) and \ref{instant-up5}(o) still have phase singularities. 
After a very long time of evolution at $t=2000$, the system relaxes into an almost uniform state where the component densities 
and phases both are quasi-uniform as in Fig. \ref{instant-up5}(d,h) and \ref{instant-up5}(l,p), respectively.

\begin{figure}[ht]
  \includegraphics[width=\linewidth]{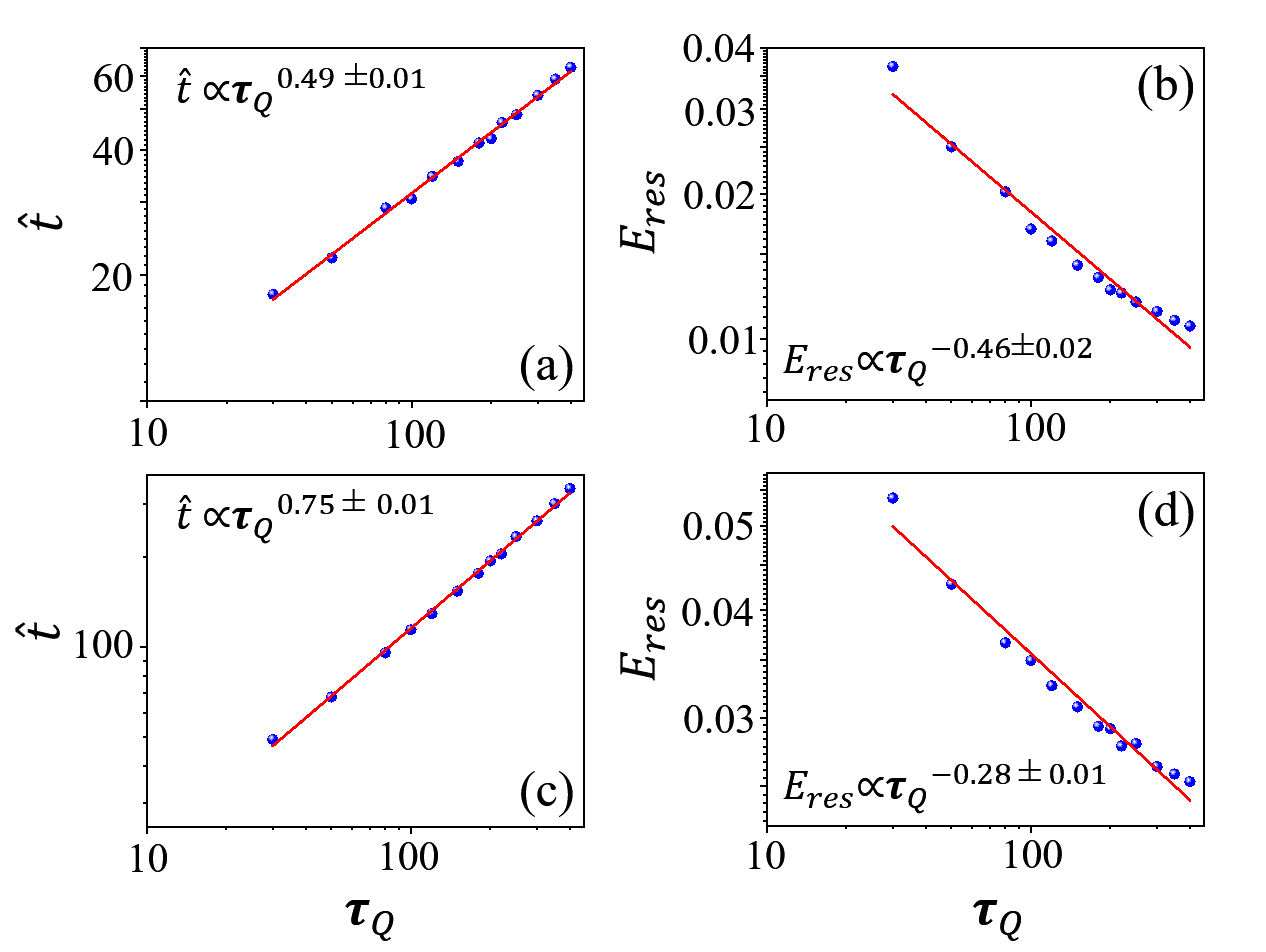}
  \caption{For MI(2)-SF transition with $U_{\uparrow\downarrow} = 0.5$: (a) $\hat{t}$ as a function of $\tau_Q$ 
  on a log-log scale with the critical exponent of $0.49\pm 0.01$ and (b) $E_{\rm res}$ as a function of $\tau_Q$
  on the log-log scale with $-0.46 \pm 0.02$ as the critical exponent. (c) and (d) are, similarly, the $\hat{t}$ 
  and $E_{\rm res}$ as functions of  $\tau_Q$ on the log-log scale
  for MI(1)-SF transition with $U_{\uparrow\downarrow} = 0.5$.} 
\label{scaling-laws-sc}
\end{figure}

To study the scaling laws, we consider a range of quench times from $\tau_Q = 30$ to $\tau_Q=400$. We measured
$\hat{t}$ corresponding to each $\tau_Q$ following the overlap protocol. $\hat{t}$ increases with an increase
in $\tau_Q$ as evident in Fig.~\ref{scaling-laws-sc}(a). On the other hand, the residual energy $E_{\rm res}$ 
decreases with $\tau_Q$ as reported in Fig.~\ref{scaling-laws-sc}(b) as $J(\hat t)$ approaches 
the critical tunneling strength $J_c$ with an increasing $\tau_Q$, [cf. Table-\ref{jval}]. Both observables 
follow power-law scaling with the critical exponents $\nu=0.45$ and 
$z=2.13$. It is pertinent to note that the critical values obtained in our scaling analysis are in very close
agreement with the values predicted by the mean-field theory 
($\nu = 0.5$ and $z = 2$).

\begin{table}[h]
\caption{The $J(\hat{t})$ for different $\tau_Q$'s during quench dynamics from MI(2) to SF phase
with $\mu = 1$ and $U_{\uparrow\downarrow} = 0.5$. The critical value of hopping strength 
$J_{\rm c}=0.042$.}
  \begin{tabular}{|c||c|c|c|c|c|c|c|c|} \hline
    $\tau_Q$     & $30$    & $50$    & $80$    & $100$   & $150$   & $200$   & 
    $300$   & $400$   \\ \hline
    $J(\hat{t})$ & $0.055$ & $0.052$ & $0.050$ & $0.049$ & $0.048$ & $0.047$ & $0.046$ & $0.045$\\ \hline
  \end{tabular}
\label{jval}
\end{table}
\subsubsection{MI(1) to SF phase transition}
\label{dynamicsB-2}
The key difference between MI(2) and MI(1) of TBHM is the density in-homogeneity as shown in the equilibrium 
lattice-site distributions [see Figs.~\ref{FIG1}(e)-(h)]. 
This results in a number of differences in quench 
dynamics. One is that the impulse regime size increases  compared to MI(2)-SF transition for each $\tau_Q$. 
This is supported by the fact that, for MI(2)-SF transition $J(\hat{t})$ always lies between $J_c$ to $2J_c - J_i$, 
but here it is not the case. Fixing $\mu$ at $0.25$, we quench $J$ from $J_i=0.01$ to a sufficiently high hopping
strength $J_f = 0.0604$, so that $J(\hat{t})$ lies in between $J_i$ and $J_f$. The critical value $J_c$ of MI(1)-SF
transition is $0.0268$. For smaller $\tau_Q$'s up to about $\tau_Q=150$, $J(\hat{t})$ is greater than 
$2J_c - J_i = 0.0436$. Even for the largest value of $\tau_Q=400$ considered in the present study, $J(\hat{t})$ 
is $0.041$ as reported in Table~\ref{jval2}.
\begin{table}[h]
\caption{The $J(\hat{t})$ for different $\tau_Q$'s during the quench dynamics from
MI(1) to SF phase with $\mu = 0.25$ and $U_{\uparrow\downarrow} = 0.5$.
The critical hopping strength is $J_{\rm c} = 0.0268$.}
  \begin{tabular}{|c||c|c|c|c|c|c|c|c|} \hline
    $\tau_Q$     & $30$    & $50$   & $80$    & $100$   & $150$   & $200$   & 
    $300$   & $400$   \\ \hline
    $J(\hat{t})$ & $0.054$ & $0.05$ & $0.047$ & $0.046$ & $0.044$ & $0.043$ & $0.042$ & $0.041$\\ \hline
  \end{tabular}
\label{jval2}
\end{table}
\begin{figure}[ht]
  \includegraphics[width=0.8\linewidth]{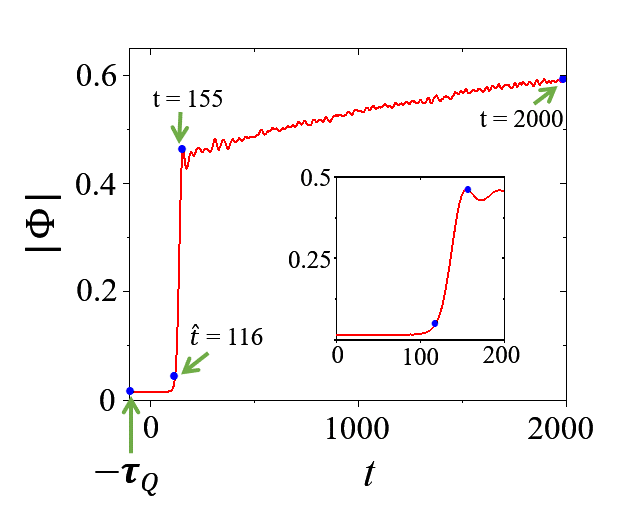}
  \caption{$|\Phi|$ as a function of time $t$ for MI(1)-SF transitions at $U_{\uparrow\downarrow}=0.5$, $\mu = 0.25$
  and $\tau_Q = 100$. Blue points are the temporal markers referred to in the main text. $|\Phi|$ remains nearly zero 
  for $t<\hat{t} = 116$. The hopping quench is performed until $t=130$. The exponential increase in $|\Phi|$ followed 
  by oscillations are shown in the inset with $t$ varying from $t=0$ to $t=200$.} 
\label{phi-up5}
\end{figure}
Another difference is that due to the absence of particles of at least one species 
at each lattice site, vortex density does not give a fair idea about the actual number of vortices. 
Due to the increase in transition time compared to the MI(2)-SF case, 
the exponent of transition time with $\tau_Q$ is higher, whereas the exponent of residual energy is lower. 
However, transition time and residual energy still follow power-law scaling with $\tau_Q$ as shown in Figs.~\ref{scaling-laws-sc}
(c) and (d).
\begin{figure}[ht]
  \includegraphics[width=\linewidth]{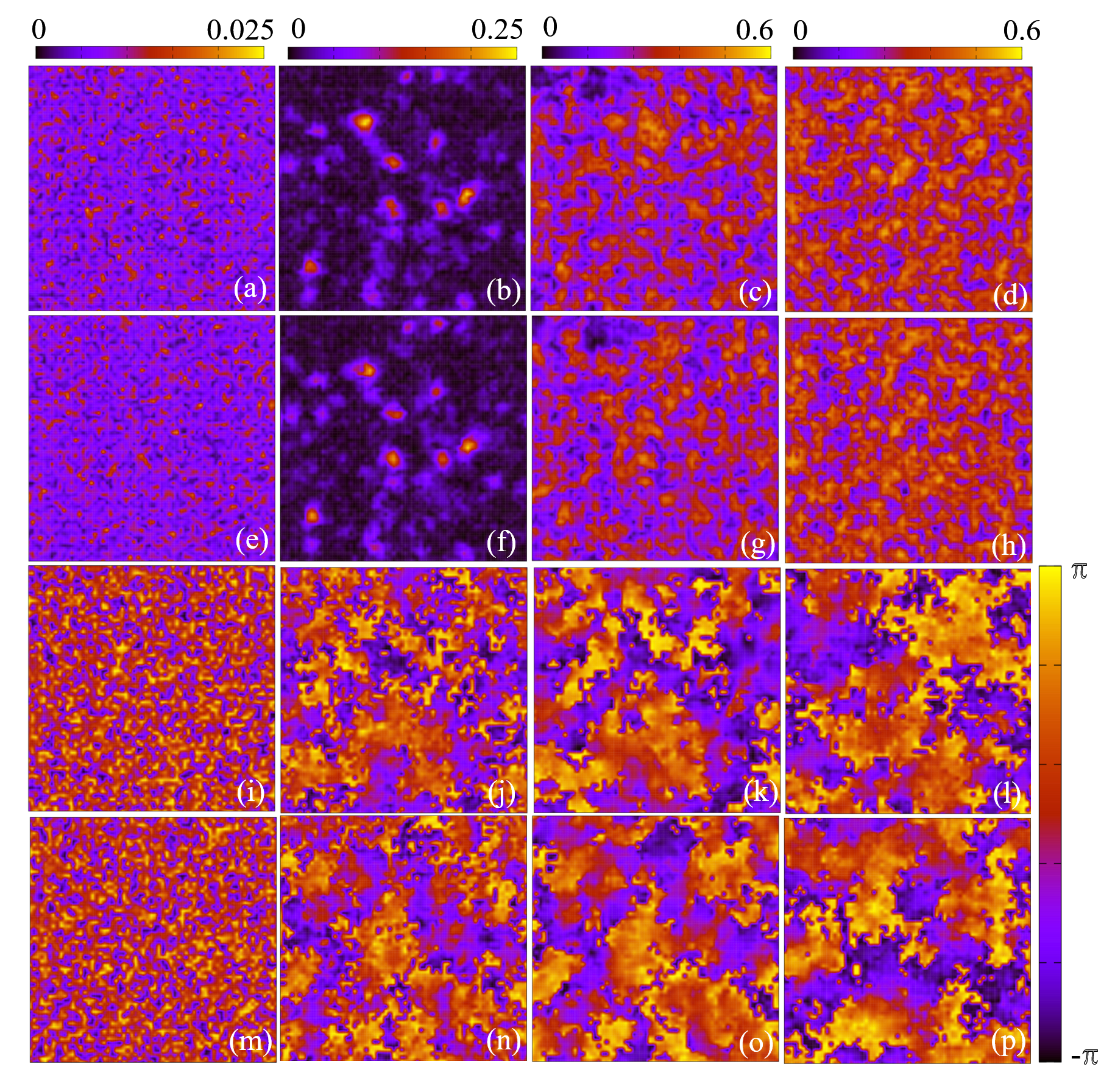}
  \caption{
  $|\phi^\uparrow_{p,q}|$ at (a) $t=-100$, (b) $t= 116$, (c) $t = 155$, and (d) $t = 2000$ for MI(1)-SF
  phase transitions corresponding to the time evolution in Fig. \ref{phi-up5}. Similarly, $|\phi^\downarrow_{p,q}|$ 
  at the same instants are in (e)-(h). Phases corresponding to $|\phi^\uparrow_{p,q}|$ in (a)-(d) are shown in (i)-(l),
  and the same for  $|\phi^\downarrow_{p,q}|$ in (e)-(h) are in (m)-(p).} 
\label{D-u12-p5-mi1}
\end{figure}

We have shown the evolution of the superfluid order parameter for $\tau_Q=100$ in Fig.~\ref{phi-up5}.
Quench is stopped at $t= 130$ in this case. $|\Phi|$ is close to zero till $t=\hat{t} = 116$, after which it increases rapidly 
followed by oscillations persisting over long periods during which it increases gradually. This is in stark contrast to 
the MI(2)-SF transition, where the order parameter stabilizes over a longer time. 
To see how differently the state relaxes after the quench, we provide the snapshots of the amplitude and corresponding phases
of $\phi^\sigma_{p,q}$ at various time instants in Fig.~\ref{D-u12-p5-mi1}.  
Randomized $|\phi^\uparrow_{p,q}|$, $|\phi^\downarrow_{p,q}|$ and the
corresponding phases at the start of the quench at $t = -100$ are shown in Figs.~\ref{D-u12-p5-mi1} (a), (e), (i), and (m), 
respectively. Next at $t = \hat{t} = 116$, these quantities are shown in the second column of Fig.~\ref{D-u12-p5-mi1},
where a few superfluid domains have started to appear. The number of superfluid domains has increased and phases of 
the order parameters also exhibit domain formation in the third column at $t=155$; this is when oscillations in |$\Phi$| are 
triggered. Even after a long period of evolution at $t=2000$, the $\phi^\sigma_{p,q}$'s do not achieve homogeneous 
distributions, unlike for MI(2)-SF transition.

\subsection{$U_{\uparrow\downarrow} = 0.9$}
\label{dynamicsC}
\begin{figure}[ht]
  \includegraphics[width=\linewidth]{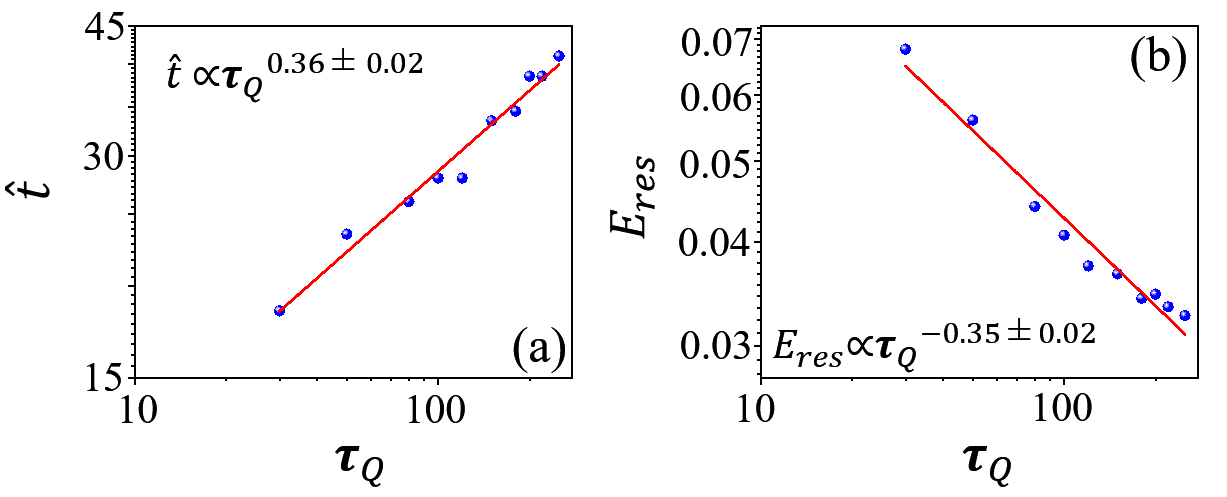}
  \caption{For first order MI(2)-SF phase transition at $U_{\uparrow\downarrow} = 0.9$ and $\mu = 1.33$: (a) $\hat{t}$ as a function of $\tau_Q$ 
  with the critical exponent of  $0.36\pm 0.02$ and (b) $E_{\rm res}$ as a function of $\tau_Q$ on log-log scale with $-0.35 \pm 0.02$ as the critical
  exponent.} 
\label{sc-laws-up9-mi2}
\end{figure}

We further discuss quench dynamics at higher interspecies interaction strength  
close to the immiscibility criterion but still in the miscible domain. 
The phase transition for $\mu = 1.33$ is first order in nature as compared to the MI(2)-SF 
and MI(1)-SF for $U_{\uparrow\downarrow}=0.5$. 
Although some traits of second-order MI(2)-SF transition, discussed in Sec. \ref{dynamicsB-1} 
like slowing down of the transition, oscillations of $|\Phi|$ about a fixed value, order parameter 
relaxing into an almost complete uniform state after free evolution, and power law scaling of $\hat{t}$ 
and $E_{\rm res}$ with $\tau_Q$ are present, the striking difference appears in the exponents in Fig.~\ref{sc-laws-up9-mi2}.   

The quench dynamics and the exponents in the scaling laws of second-order MI(1)-SF transition 
for $U_{\uparrow\downarrow} = 0.9$ remain almost similar to MI(1)-SF transition for $U_{\uparrow\downarrow} = 0.5$ 
as discussed in Sec.~\ref{dynamicsB-2} and are not shown here for brevity.
\subsection{$U_{\uparrow\downarrow} = 1.5$}
\label{dynamicsD}
Finally, we discuss quench dynamics for the phase-separated regime, 
where one of the two components occupies the lattice site. 
\begin{figure}[ht]
  \includegraphics[width=0.8\linewidth]{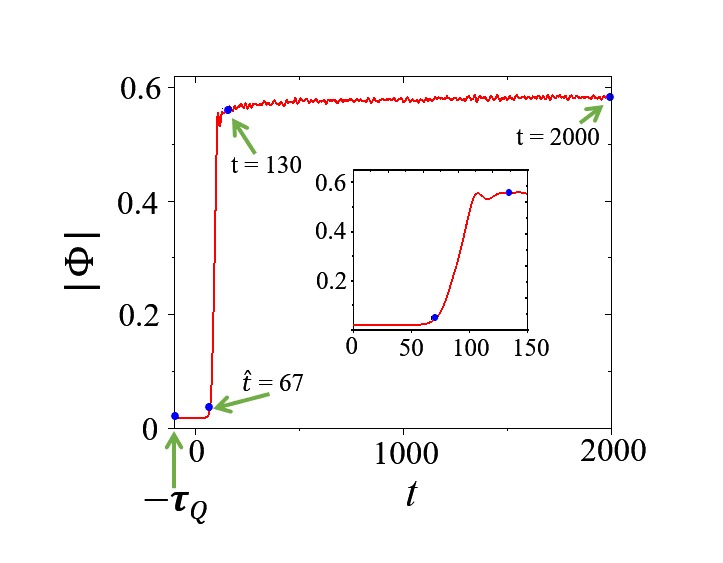}
  \caption{The evolution of $|\Phi|$ with $t$ for MI(2)-SF transitions at $U_{\uparrow\downarrow} = 1.5$, $\mu = 1.45$ and $\tau_Q=100$. 
  Blue points indicate the temporal markers referred to in the main text. $|\Phi|$ is close to zero until $\hat{t}=67$. 
  The exponential increase in $|\Phi|$ followed by oscillations are shown in the inset with $t$ varying from $t=0$ to $t=150$.} 
\label{plot-phi-1}
\end{figure}
Starting from $\mu=1.45$ and $J=0$ which corresponds to the MI(2) lobe, we perform a quench that terminates at $J=0.0516$, lying well within the SF phase.
 The transition is indicated by the growth of the superfluid order parameter and occurs after crossing the critical $J_c=0.0258$, as demonstrated in
 Fig.~\ref{plot-phi-1} for $\tau_Q = 100$. $|\Phi|$ remains close to zero until $t= \hat{t} = 67$, followed
 by a rapid increase period. 

The quench is halted at $t = \tau_Q = 100$, but we freely evolve the system up to $t=2000$ to confirm the ground state of the system. 
\begin{figure}[ht]
  \includegraphics[width=\linewidth]{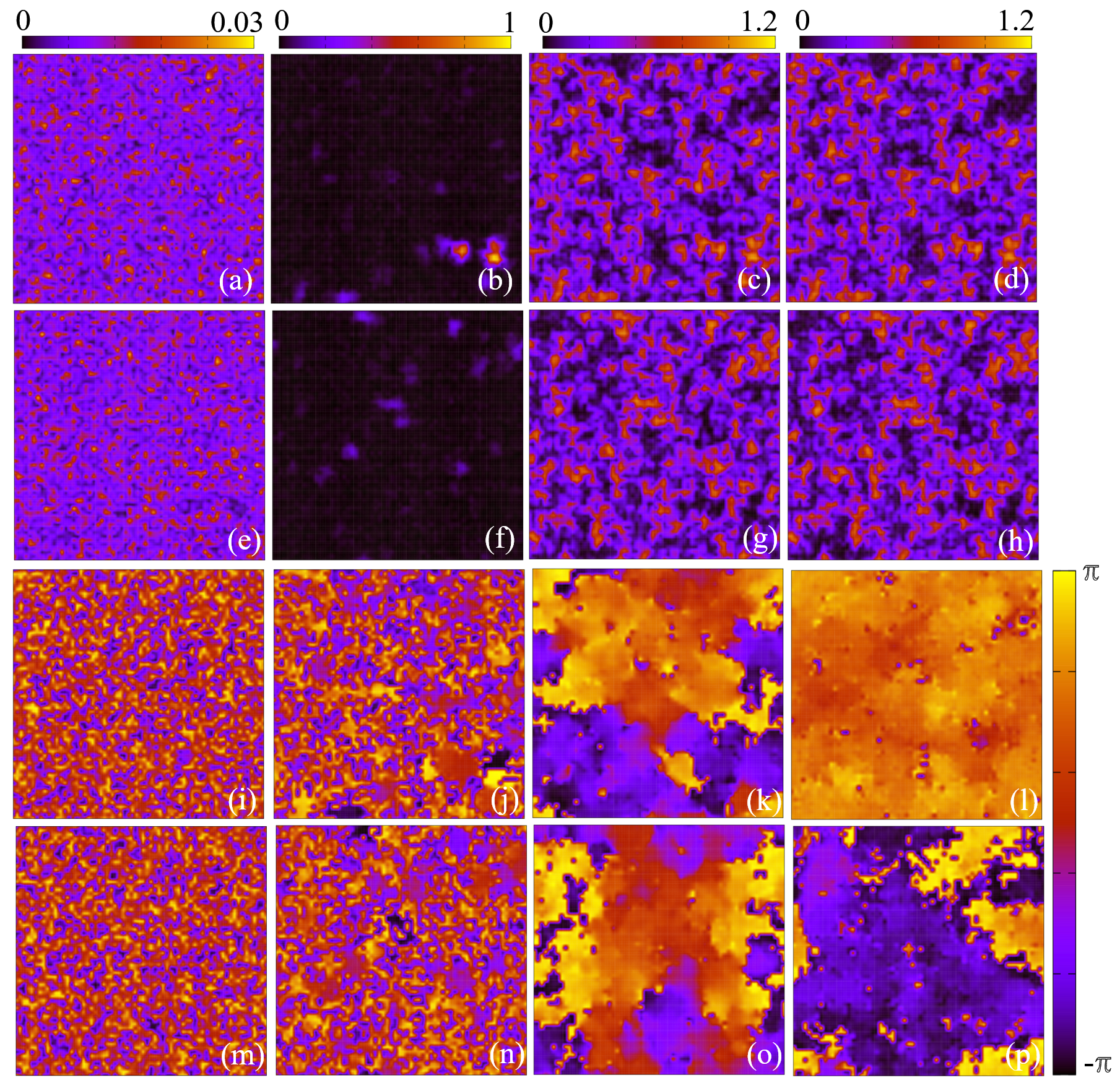}
  \caption{
   $|\phi^\uparrow_{p,q}|$ at (a) $t=-100$, (b) $t= 67$, (c) $t = 130$, and (d) $t = 2000$ for MI(2)-SF
  phase transitions corresponding to the time evolution in Fig. \ref{plot-phi-1}. Similarly, $|\phi^\downarrow_{p,q}|$ 
  at the same instants are in (e)-(h). Phases corresponding to $|\phi^\uparrow_{p,q}|$ in (a)-(d) are shown in (i)-(l),
  and the same for  $|\phi^\downarrow_{p,q}|$ in (e)-(h) are in (m)-(p).} 
\label{DP1&2-U12-1P5}
\end{figure}
The Figs.~\ref{DP1&2-U12-1P5}(a)-(p) provide snapshots of $|\phi^{\sigma}_{p,q}|$ and the corresponding phase of $\phi^{\sigma}_{p,q}$ at 
different times. At $t=\hat{t} = 67$, domain formation begins as indicated in \ref{DP1&2-U12-1P5}(b) and \ref{DP1&2-U12-1P5}(f); the 
respective phases are shown in \ref{DP1&2-U12-1P5}(j) and \ref{DP1&2-U12-1P5}(n). At $t = 130$, both $|\phi^\uparrow_{p,q}|$ and 
$|\phi^\downarrow_{p,q}|$ have acquired many domains, as indicated in \ref{DP1&2-U12-1P5}(c) and \ref{DP1&2-U12-1P5}(g), while the 
respective phases in \ref{DP1&2-U12-1P5}(k) and \ref{DP1&2-U12-1P5}(o) start exhibiting domain merging. 
The component order parameters at $t=2000$ show no discernible difference from those at $t=130$, as shown in \ref{DP1&2-U12-1P5}(d), 
(h), and \ref{DP1&2-U12-1P5}(l), (p).
\begin{figure}[ht]
  \includegraphics[width=\linewidth]{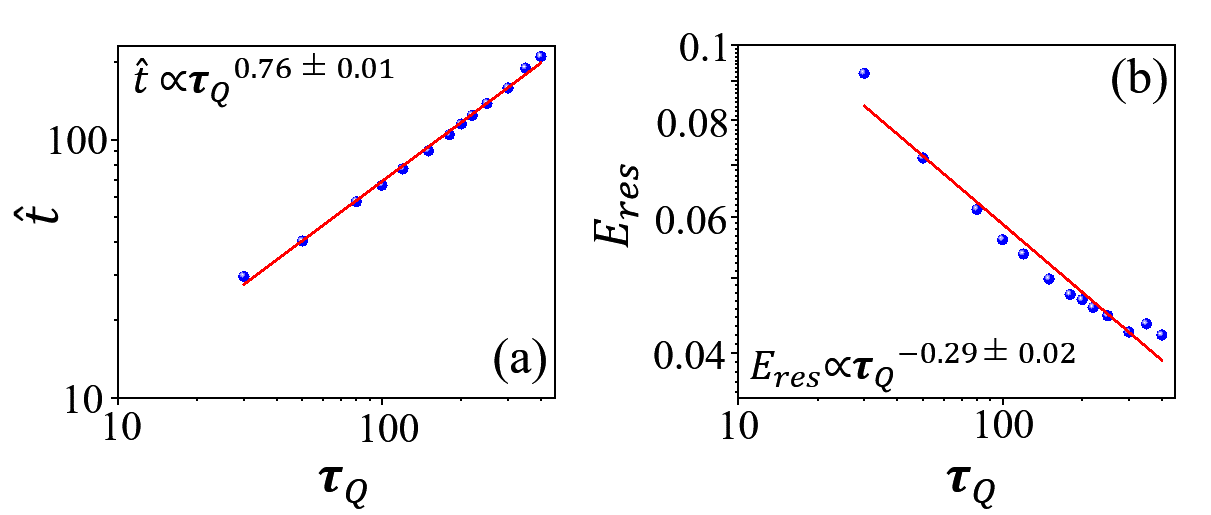}
  \caption{For MI(2)-SF phase transition at $U_{\uparrow\downarrow} = 1.5$: (a) $\hat{t}$ as a function of $\tau_Q$ 
  with the critical exponent of $0.76\pm 0.01$ and
  (b) $E_{\rm res}$ as a function of $\tau_Q$ on log-log scale
  with $-0.29 \pm 0.01$ as the critical exponents. The critical exponents are similar to those displayed in Figs.~\ref{scaling-laws-sc}(c)-(d).} 
\label{scaling-laws-u1p5-mi2}
\end{figure}

Likewise, the dynamics at the other two $U_{\uparrow\downarrow}$ values discussed previously, 
$\hat{t}$ and $E_{\rm res}$ follow power-law scaling as shown in Figs.~\ref{scaling-laws-u1p5-mi2}(a)-(b) and with
the critical exponents similar to those for MI(1)-SF transition in the miscible domain. This is evident from 
Figs.~\ref{scaling-laws-sc}(c)-(d) and Figs.~\ref{scaling-laws-u1p5-mi2}(a)-(b). Since the system is in the immiscible phase, 
the nature of MI(1) and MI(2) is similar. This is concluded while studying the static properties of these systems. The quench dynamics of
MI(1)-SF for $U_{\uparrow\downarrow}=1.5$ is similar to that of MI(2)-SF transition, and therefore not presented here.

\section{Conclusions}
\label{summary}
We have studied the out-of-equilibrium quench dynamics of the two-component 
Bose-Hubbard model. The equilibrium phase diagram and related phase transitions depend 
on interspecies interaction strength. We performed the quench 
of the hopping strength across the MI-SF phase transitions and observed 
that the average filling of the MI lobes and the order of the phase transitions 
lead to different dynamics from what is observed in a single-component BHM. The MI-SF phase transitions, 
in the miscible regime, from the Mott lobes with average occupancy $1$ or $2$ are 
second-order for $U_{\uparrow\downarrow}$ less than a critical strength above which 
the transitions at and around the tip of the MI(2) lobe are first-order in nature. 
The critical exponents calculated through scaling analysis for the
second-order phase transition from the Mott lobes with homogeneous atomic occupancy distributions
to the superfluid phase are in good agreement with the mean-field predictions. 
Due to the atomic occupancy inhomogeneity in MI(1), the impulse regime is 
extended across MI(1)-SF phase transition at $U_{\uparrow\downarrow}=0.5$. 
The quench dynamics also reveal that the order of the transitions greatly affects 
the critical exponents. In the immiscible regime, the power-law scaling of the exponents is 
maintained with exponents similar to those for MI(1)-SF phase transition in the miscible regime. 
We hope that the phenomena discussed in the present work could be realized in cold-atom experiments 
on strongly-correlated bosonic mixtures in optical lattices. Our study serves as a route to understand 
the dynamics of quantum phase transitions in spin-orbit coupled condensates in optical lattices. 
The exploration in this direction may unveil the role of coupling in quantum mixtures and the applicability 
of the Kibble-Zurek mechanism to two-component Bose-Hubbard models.

\begin{acknowledgments}
  We thank Deepak Gaur and Hrushikesh Sable for valuable discussions. K.S. acknowledges the support from IAMS, 
  Academia Sinica, Taiwan. S.G. acknowledges support from the Science and Engineering Research Board, Department
  of Science and Technology, Government of India through Project No. CRG/2021/002597.
\end{acknowledgments}

\end{document}